\documentclass{article}

\usepackage[final]{neurips_2024}

\usepackage[utf8]{inputenc} % allow utf-8 input
\usepackage[T1]{fontenc}    % use 8-bit T1 fonts       % hyperlinks
\usepackage{url}            % simple URL typesetting
\usepackage{booktabs}       % professional-quality tables
\usepackage{amsfonts}       % blackboard math symbols
\usepackage{nicefrac}       % compact symbols for 1/2, etc.
\usepackage{microtype}      % microtypography
\usepackage{graphicx}
\usepackage{caption}
\usepackage{cite}
\usepackage{amsmath}
\usepackage{subcaption}
\usepackage{placeins}
\usepackage{natbib}
\setcitestyle{numbers}
\usepackage{hyperref}
\usepackage{xcolor}

\title{Do LLM Personas Dream of Bull Markets? Comparing Human and AI Investment Strategies Through the Lens of the Five-Factor Model}

\author{%
  Harris Borman \\
  Commonwealth Bank of Australia \\
  \texttt{harris.borman@cba.com.au} \\
   \And
  Anna Leontjeva \\
  Commonwealth Bank of Australia \\
  \texttt{anna.leontjeva@cba.com.au} \\
   \AND
   Luiz Pizzato \\
   Commonwealth Bank of Australia \\
   \texttt{luiz.pizzato1@cba.com.au} \\
   \And
    Max Kun Jiang \\
    Commonwealth Bank of Australia \\
   \texttt{max.jiang@cba.com.au} \\
   \And
   Dan Jermyn \\
   Commonwealth Bank of Australia \\
   \texttt{dan.jermyn@cba.com.au} \\
}

\begin{document}

\maketitle

\begin{abstract}
Large Language Models (LLMs) have demonstrated the ability to adopt a personality and behave in a human-like manner. There is a large body of research that investigates the behavioural impacts of personality in less obvious areas such as investment attitudes or creative decision making. In this study, we investigated whether an LLM persona with a specific Big Five personality profile would perform an investment task similarly to a human with the same personality traits. We used a simulated investment task to determine if these results could be generalised into actual behaviours. In this simulated environment, our results show these personas produced meaningful behavioural differences in all assessed categories, with these behaviours generally being consistent with expectations derived from human research. We found that LLMs are able to generalise traits into expected behaviours in three areas: learning style, impulsivity and risk appetite while environmental attitudes could not be accurately represented. In addition, we showed that LLMs produce behaviour that is more reflective of human behaviour in a simulation environment compared to a survey environment.
\end{abstract}

\section{Introduction}

Large Language Models (LLMs) have demonstrated the ability to adopt human personas to produce a believable simulation of human behaviour \citep{park2023}. Past works have investigated LLM powered simulations of human personalities \citep{winter2024} and the effect of these personalities on model output \citep{jiang2024}. There has been some research into the downstream behavioural impacts of these personalities in simulations \citep{noh2024}. However, these works are limited as they primarily focus on whether different behaviours can be produced through assigned personalities, rather than examining if these behaviours are truly representative of a human population.

For LLM-powered simulations to be generally applicable to business problems, they need to accurately represent a broad range of human behaviours. If they can only reliably represent a small subset of personalities, any results will inherently be biased towards those groups. Therefore, any reliance on these systems for any business activities will exhibit the same biases, and potentially discriminate against other groups. This study aims to address this limitation by investigating if LLM-powered personas \footnote{We will refer to an LLM-powered persona as simply "persona" for the remainder of the paper} can reliably interpret a human personality model (specifically the five-factor model) and map personality traits into specific behaviours that are consistent with past human research. We will do this by investigating the consistency and persistance of simulated behaviours in investment-related decision-making. By doing so, we aim to assess whether the simulated traits produce coherent behavioural patterns across diverse scenarios, similar to the relationship between personality and investment related behaviours observed in a human population.

Extensive research has been conducted to determine the behavioural impacts of one’s personality. These studies have explored various behaviours including investment attitudes \citep{gambetti2019}, creative decision making \citep{saihani2009} and learning style \citep{siddiquei2018}. Various models exist as means of simplifying human personality and representing it as a combination of traits. The five-factor model \citep{mccrae1991} proposes human personality as a combination of the following traits: 
\begin{itemize}
  \item Openness: A trait that represents a need for variety, novelty and change. 
  \item Conscientiousness: A trait that represents achievement striving and aspirational behaviours. 
  \item Agreeableness: A trait that represents compliance and social deference. 
  \item Extraversion: A trait that represents companionship and social stimulation preferences. 
  \item Neuroticism: A trait that represents an individual’s emotional stability. 
\end{itemize}

This model has been shown to have consistency and replicability across different methodologies \citep{biesanz2004} and has proven validity across cultures \citep{mount1998}; hence becoming one of the most used metrics for personality assessment \citep{corr2020}. We lean on these findings and will use the five-factor model in our study.   

This study explores the intersection between LLM personality research and behavioural personality research, focusing on the following question:

RQ: Can LLMs accurately translate assigned personality traits into behaviours, specifically in investment tasks, in a manner consistent with human personality? 

To answer this, we developed a set of LLM-powered personas that encompassed a full range of human personality traits. These personas completed a short behavioural survey derived from past research to determine if they can associate personality traits with specific behaviours. The personas were then given an investment task to determine if these results could be generalised and produce meaningful behavioural differences in a simulated environment. 

\section{Methods}
\subsection{Persona development}

We built personas based on the five-factor model of personality, assigning values of low, medium, or high for each of the five personality traits. This resulted in 243 unique personas, ensuring comprehensive coverage of possible personality combinations. Personas were created with the following prompt:

{\ttfamily You are to take on the personality of the following individual\\
 Openness to Experience: (Low, Medium or High) \\
 Conscientiousness: (Low, Medium or High) \\
 Extraversion: (Low, Medium or High) \\
 Agreeableness: (Low, Medium or High) \\
 Neuroticism: (Low, Medium or High) \\}
 
The LLM was then told it would be presented with a series of questions and should respond how a human with the same personality would (See Appendix \ref{A} for a complete prompt).
 
We tested these personas using both ChatGPT 3.5 and ChatGPT 4.0. In this study, we report ChatGPT 4.0 results as it produced more reliable outcomes, although largely similar to ChatGPT 3.5 (See Appendix \ref{D} for ChatGPT 3.5 results). 
 
We validated our personas prompt by asking them to complete a five-factor model personality test. This was done to ensure that the LLM is correctly using the personality traits to answer questions and identify any prompting issues (See Appendix \ref{C} for persona personality test results).  

\subsection {Behavioural survey}

To measure if personas can correctly associate personality with behaviour, we constructed a simple behavioural survey. This survey contained 9 questions (See Appendix \ref{A} for specific questions) and was primarily created from questions used in past research \citep{gambetti2019,busicsontic2017}. By using questions that have been tested in past research there was a direct benchmark to compare against the LLM results. This allowed us to reliably quantify how well the personas are replicating human behaviours.  
 
We chose these questions to measure a range of behaviours such as learning style and investment attitudes that do not have an obvious association with any personality traits. This design served several purposes related to enhancing the robustness and validity of our study. We challenged LLM personas to make inferences between personality traits and behaviours, potentially demonstrating a deeper "understanding" of trait implications. We aimed to minimize the possibility that the LLM was simply reproducing learned correlations from training data as it appeared in \citep{gambetti2019} and \citep{busicsontic2017}.  

To further validate the LLM's ability to accurately simulate personality-driven behaviours, we implemented a second experimental stage. In this stage, personas performed a task-based simulation where their behaviour could be observed rather than described. This step allowed us to: 
\begin{itemize}
  \item Determine if the personas could consistently apply their personality traits in novel situations.
  \item Assess whether the behavioural patterns observed in the survey extended to more complex, open-ended tasks. 
  \item Identify any discrepancies between survey responses and actual behaviour in simulated scenarios. 
\end{itemize}
This approach helped us to distinguish between mere reproduction of learned correlations and genuine personality-driven behaviour simulation.

\subsection {Investment simulation}

To determine if LLMs were interpreting their personality traits to produce different behaviours, as opposed to simply learning the expected answer to each question during training, we constructed a simple investment task. This task also let us evaluate the consistency of each of these behaviours. The task was created originally and not derived from any content in existing literature to avoid any risk of the model being trained on the content. The task was designed to allow us to measure each of the behaviours tested during the survey without explicitly stating them. By doing this we aimed to force the LLM to translate its knowledge on personality traits and their associated behaviours into specific actions.
 
For this task each persona was told that it had \$1000 to invest in 1 of 5 companies. Personas could learn more about each company up to 5 times through either independent research or talking to an expert. Personas were told that they would get identical knowledge from either of these approaches, to avoid any assumptions about these methods affecting behaviour. To prevent any filler text describing the companies from influencing behaviour, personas were not given any actual information when they chose to learn about a company. Instead, they were simply told in each prompt how many times they had researched each company. Personas were given a brief summary of each company containing its name, an expected return on investment (ROI) and a risk factor, indicating the chance that they would lose their investment (See Appendix \ref{B} for example prompt). The LLM was then presented with this prompt repeatedly, at each stage outputting a company name and research method. This continued until either the persona independantly made an investment decision, or in the case that each company had been researched the maximum number of times, the LLM was directed to make an investment decision.
 
The selection of companies was as follows 
\begin{itemize} 
  \item Diamond: ROI: 5\%, Risk 10\%  
  \item Platinum: ROI 35\%, Risk 30\% 
  \item Emerald: ROI 89\%, Risk 50\%  
  \item Ruby: ROI 25\%, Risk 30\% (An eco-conscious company) 
  \item Sapphire: ROI 80\%, Risk 60\% (A cutting edge technology company)
\end{itemize}

The first 3 companies were designed so each investment would have the same expected value\footnote{
\[
EV = \sum_{i=1}^{n} V_i \cdot P_i 
\]

Where:
\begin{itemize}
  \item $EV$ = Expected Value of Investment
  \item $V_i$ = Potential Investment Value 
  \item $P_i$ = Probability of Achieving $i$th Investment Value
  \item $n$ = Number of Possible Outcomes
\end{itemize}
}. We did this to prevent the LLMs from simply selecting the investment with the highest expected value. By giving each company the same expected value, the LLM was forced to simulate the personas risk appetite when making their decision. The final 2 companies were designed to be worse investments financially. This was done to identify values that personas would place higher than financial return. For these companies personas were also given a brief description of the company, so this could be factored into the decision-making process. 

During the simulation we collected data on which companies were investigated at each stage, the number of times each company was investigated, the method of investigation chosen at each step and the personas final investment decision. These data points all gave insights into specific behaviours tested in the behavioural survey. For both tests we used an ordinary least squares (OLS) regression to identify correlations between personality traits and each desired behaviour.  

\section{Results}
\subsection{Learning style}

\begin{figure}
  \centering
 
\includegraphics[width=0.8\textwidth]{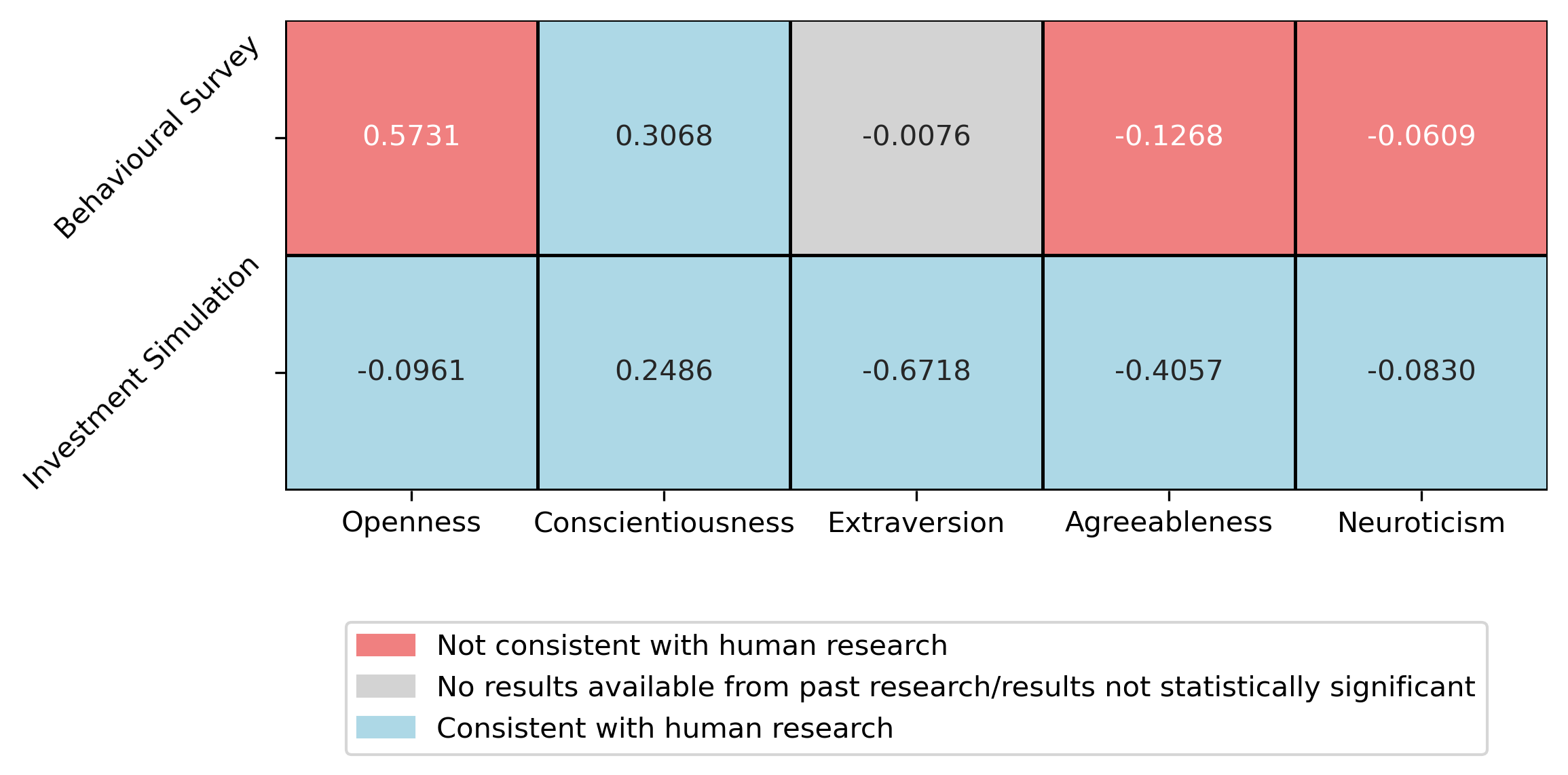}
\caption{Relationship between personality traits and learning style}
\label{Figure1}
\end{figure}

Learning style was assessed in the behavioural survey by a single question measuring whether a persona would prefer to learn through independent research or by talking to a subject matter expert. In the simulation, this behaviour was measured by the time the persona spent researching a company using these methods. In our results, a positive correlation represented a preference for learning independently, while a negative correlation represented a preference for learning from others.

Past research gives us the expectation that openness and extraversion should both be negatively correlated with a reflective learning style \citep{siddiquei2018}, so we expect personas with high values for these traits to learn from others. Other research focused on Kolb’s learning style model found correlations between neuroticism \citep{kamarulzaman2012} and agreeableness \citep{akbar2020} and an accommodating learning style, characterized by a preference to rely on others for information. Conscientiousness has been observed to have a positive correlation with an assimilating learning style \citep{kamarulzaman2012}. This learning style is characterized by learning through readings and exploring analytical models, work that is generally done individually. 

In the behavioural survey persona behaviour was somewhat consistent with these findings producing the correct behaviours for conscientiousness, agreeableness and neuroticism (see Figure \ref{Figure1}).  Openness was the only trait where the behaviour did not align with past research and no statistically significant correlation could be drawn for extraversion. In the simulation, personas were able to accurately represent this facet of human behaviour producing the expected behaviour for all 5 traits. This shows that LLM personas offer more accurate results when interpreting personality traits into learning style behaviour in a simulation environment compared to a survey environment. 

\subsection{Impulsive decision making}

\begin{figure}
  \centering
 
\includegraphics[width=0.8\textwidth]{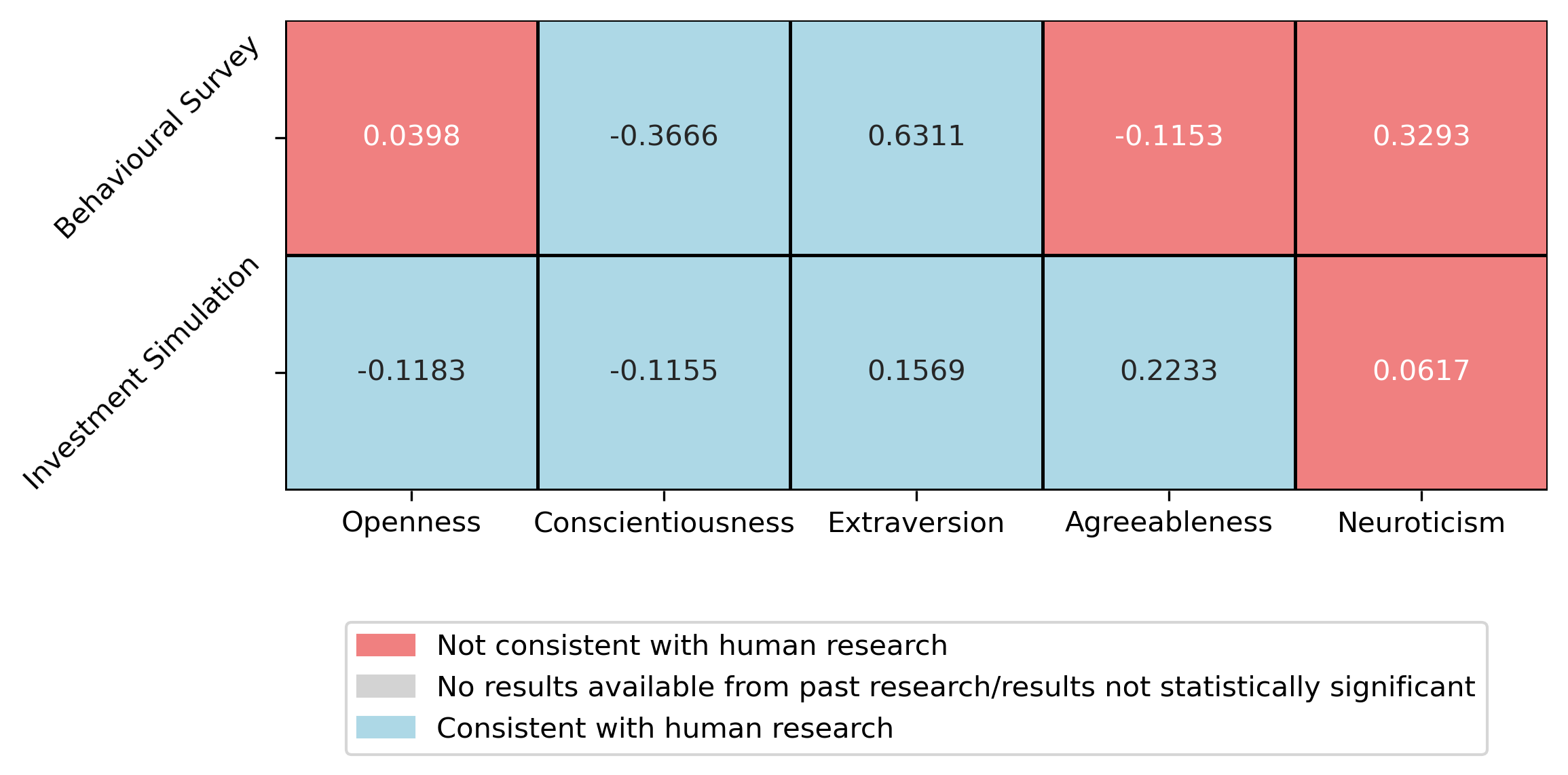}
\caption{Relationship between personality traits and impulsive decision making}
\label{Figure2}
\end{figure}

In the behavioural survey, impulsive decision making was calculated using 2 questions measuring impulsive decision making and instinctive decision making. In the simulation, this behaviour was measured by the time a persona spent researching before making an investment decision. 

Past research gives us the expectation that extraversion should be correlated with impulsive decision making and reduced planning time while neuroticism should be negatively correlated with this behaviour \citep{weinman1984}. Conscientiousness has been observed to show a negative correlation with impulsivity in areas including a lack of perseverance, lack of premeditation and a sense of urgency \citep{whiteside2001}. As such, we expect conscientiousness to be negatively correlated with impulsive decision making. Agreeableness on the other hand has demonstrated a positive correlation with impulsive spending \citep{turkyilmaz2014}. Openness has shown some correlations with the sensation seeking aspect of risk taking \citep{whiteside2001}. This aspect, however, is characterized only by seeking new and varied experiences that manifests in risk taking behaviour. Additionally, since our measurement of impulsive decision-making is constructed by a persona’s desire to seek new knowledge on each company, we expect to see a negative correlation for openness. 
  
In the behavioural survey inconsistent results were produced. Personas were able to correctly represent the traits of conscientiousness and extraversion but incorrectly represented openness, agreeableness and neuroticism (see Figure \ref{Figure2}). Personas achieved better results in the simulation producing the correct behaviours for openness, conscientiousness, extraversion and agreeableness. The only trait incorrectly represented in the final simulation was neuroticism. For example, while human studies give us the expectation that an individual with high neuroticism would spend more time researching before making an investment decision. In the simulation the majority of personas with high neuroticism researched a company less than 5 times out of a possible 25.
   
\subsection{Risk appetite}
\begin{figure}
  \centering
 
\includegraphics[width=0.8\textwidth]{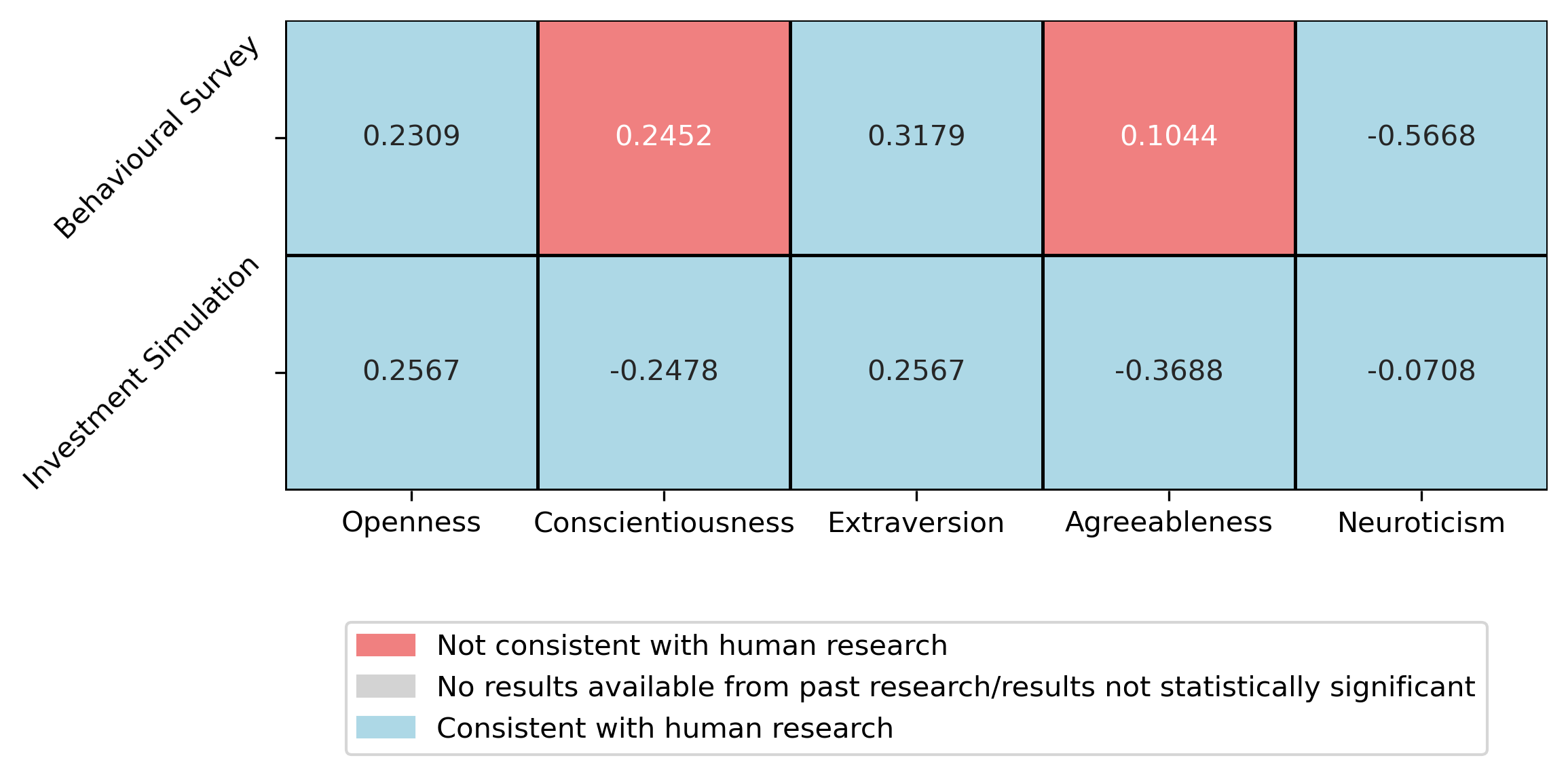}
\caption{Relationship between personality traits and risk appetite}
\label{Figure3}
\end{figure}

In the behavioural survey risk appetite was assessed with 2 questions that focused on risk and profit perception. In the simulation this behaviour was measured using the risk factor of the persona’s final investment. We defined a risky investment as an investment in one of the 2 riskiest companies, Emerald or Sapphire. In Figure 3, a positive correlation represents a more relaxed risk appetite while a negative correlation represents risk averse behaviour. 
  
Past research shows that openness is positively correlated with risk taking in relation to a financial gain while neuroticism is negatively correlated with this behaviour \citep{lauriola2001}. It also shows that both agreeableness \citep{mehregan2018, nicholson2005} and conscientiousness \citep{nicholson2005} are positively correlated with risk averse behaviour. Extraversion has also demonstrated a positive correlation with both riskier trading strategies \citep{mayfield2008} and a more relaxed risk appetite \citep{mehregan2018}. 
  
In the behavioural survey, personas were able to accurately represent the traits of openness, extraversion and neuroticism but produced inaccurate results for the traits of conscientiousness and agreeableness (see Figure \ref{Figure3}). The simulation results were able to correctly represent all 5 traits with personas constructing their portfolios in the same way as a person of the same personality would be expected to in all cases.

\subsection{Interest in environmental products}
\begin{figure}
  \centering
 
\includegraphics[width=0.8\textwidth]{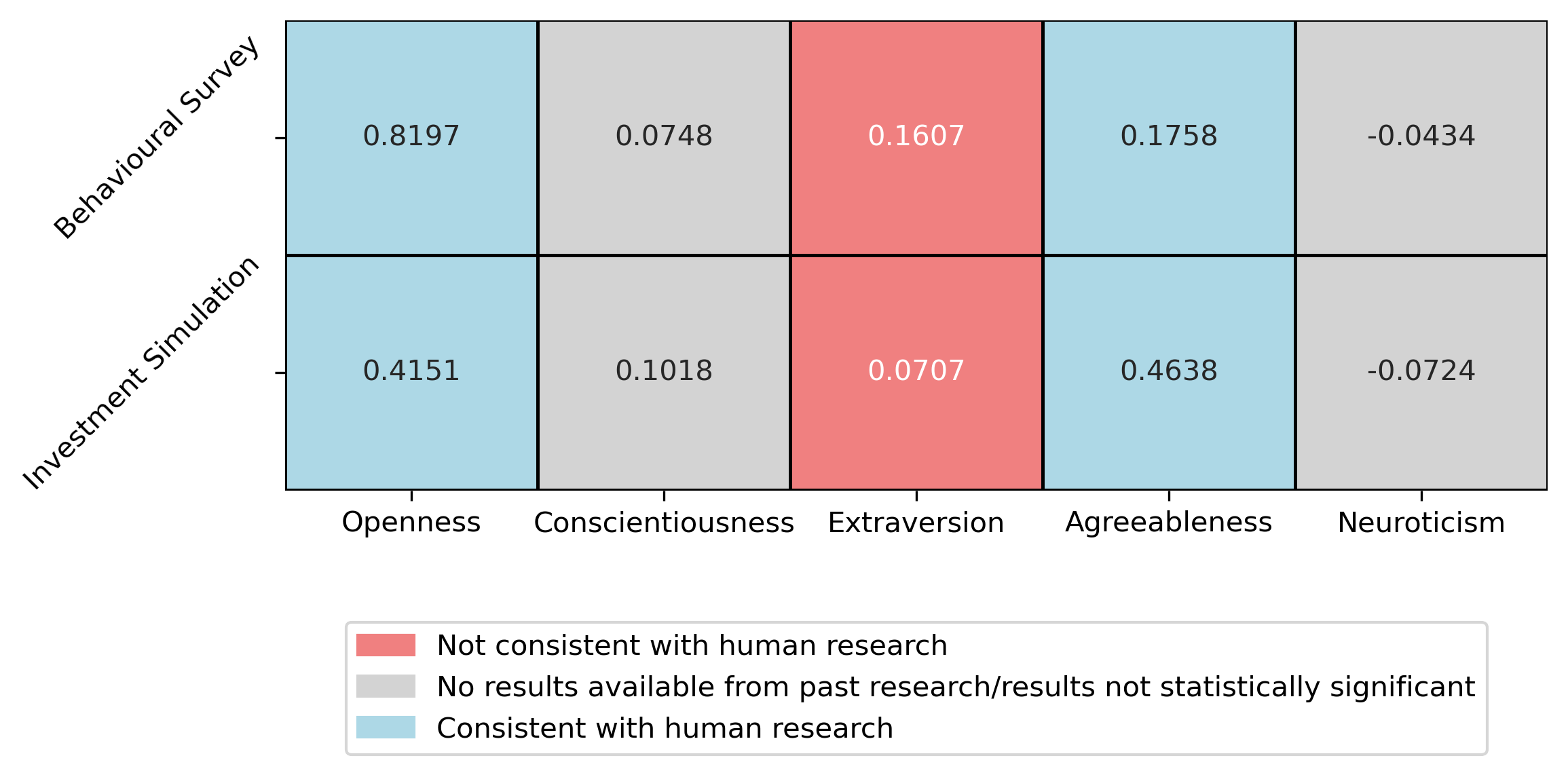}
\caption{Relationship between personality traits and an interest in environmental products}
\label{Figure4}
\end{figure}

In the behavioural survey a persona’s interest in environmental products was tested with 3 questions assessing intention/consideration to install 3 different renewable energy products (solar panels, solar water heating, wind turbines) (See Appendix \ref{A}). In the simulation, interest in environmental products was measured by the number of times a persona researched the eco-conscious company, Ruby.

Past research shows that agreeableness and openness should both be positively correlated with this behaviour while extraversion should be negatively correlated with this behaviour \citep{busicsontic2017}.

In both the behavioural survey and the investment simulation personas were able to accurately represent openness and agreeableness while extraversion was incorrectly represented (see Figure \ref{Figure4}). We were not able to identify any past research with correlations between neuroticism or conscientiousness and an interest in environmental products. As such for both the survey and simulation we were unable to compare the results to a human benchmark. These results suggest that personas in both a survey and simulation environment cannot accurately associate personality traits to an interest in environmental products.

\subsection{Investment in environmental products}
\begin{figure}
  \centering
 
\includegraphics[width=0.8\textwidth]{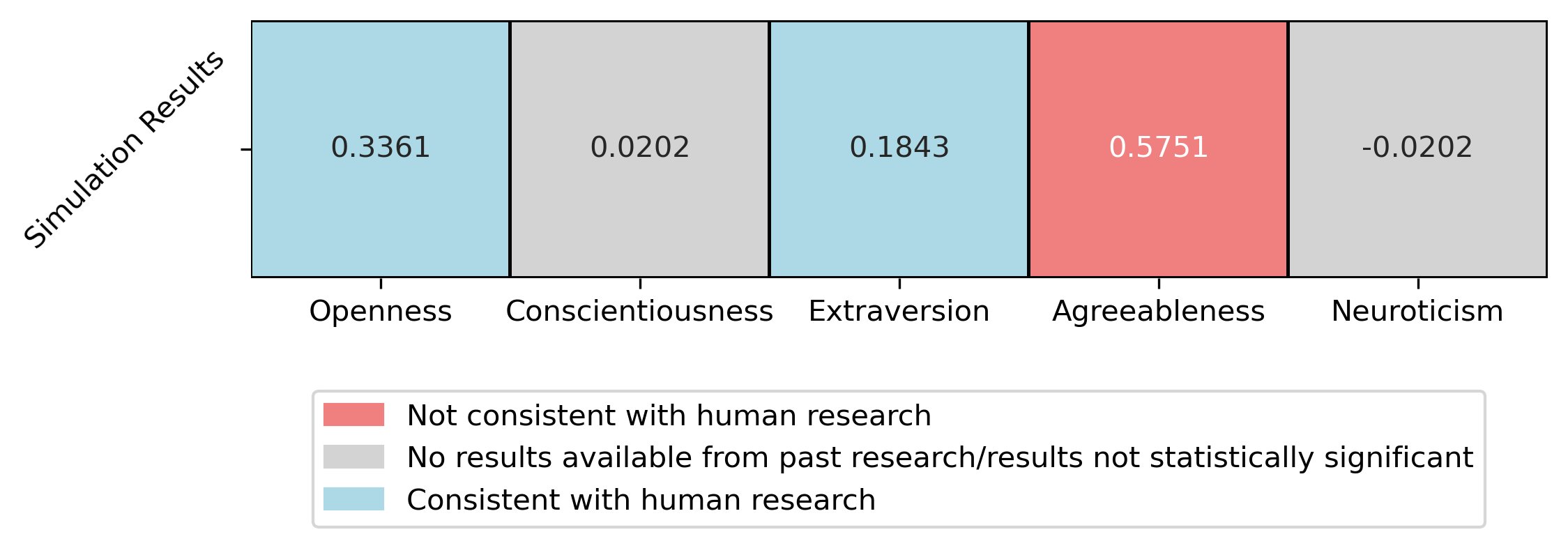}
\caption{Relationship between personality traits and investment in environmental products}
\label{Figure5}
\end{figure}

As personas in the behavioural survey were only constructed using a set of 5 trait values, investment in environmental products could not be tested without relying on the LLM to create false information. In the simulation however, this behaviour could be measured by if a persona chose to invest in the eco-conscious company Ruby. 

Past research gives the expectation that openness and extraversion should positively impact the investment decision 
while agreeableness and neuroticism should negatively impact it \citep{busicsontic2017}. 

The simulation accurately replicated behaviours for openess and agreeableness, but not for agreeableness. Additionally, no statistically significant correlation could be established for either conscientiousness or neuroticism (see Figure \ref{Figure5}).

\section{Discussion}
The results showed that in an investment scenario LLM powered personas were able to accurately reflect human behaviour in the areas of learning style, impulsive decision making, risk appetite and environmental concern in accordance with their assigned personality traits. 

It was also observed that the simulation produced a significantly better representation of human behaviour when compared to the behavioural survey results. In the simulation, personas were able to accurately represent all 5 personality traits for 2 behaviours, risk appetite and learning style. Additionally in all 4 cases where comparison was possible the simulation personas outperformed the survey personas in 3 out of 4 cases and produced similar performance in the 4th case.

This performance suggests that LLMs can simulate human behaviour better in these areas when acting in a simulated environment or operating to complete a task. This could be due to the larger amount of information present in the context window or due to the task we had the personas complete. This increased volume of information could be providing more opportunities for the LLM to create associations between behaviours and traits. Due to the uncertainty surrounding the specifics of LLM behaviour further research would be required to definitively answer the question of why this behaviour is occurring. These results suggest that LLM representation of human behaviour extends beyond learning relationships between specific questions and traits during training as this would produce significantly better performance in the behavioural survey and worse performance in the simulation. The observed behaviour suggests that during the training process the LLM has learnt to make associations between personality traits and behaviours themselves allowing them to produce an accurate representation of human behaviour in these areas when operating in a simulated environment.  

\section{Future work and limitations}
A limitation of the conducted experiments is that they are solely focused on personality traits. Prompts were constructed to have only information about the persona’s personality traits with no extra information provided (See Appendix \ref{A} for prompt). This was an intentional decision to ensure that no other factors influence the persona’s actions. In a practical implementation, more demographic information would likely be necessary to construct a comprehensive simulation. As such a promising area for further research would be to determine if these behaviours continue to persist once further information is added to the persona. 
 
Another limitation arises from controlled environment of the simulation task we used for this study. A task with a more open goal or environment could allow personas to demonstrate more subtle behavioural differences. This would come with the caveat that these behaviours may not have human research to offer a direct comparison. There has been research into the effects of personality on higher level constructs such as project success \citep{masood2018, hussain2021} that could be studied in this case. Alternatively, a future study could examine behavioural differences that arise in these tasks and evaluate them to determine if they fit within the definitions of the traits. 
 
Another area where the simulation we used is limited is the lack of inter-agent interaction. Most human research is done in a controlled environment where social interaction is not a factor, but there are contexts where it is relevant. Some traits are heavily influenced by social factors. For example, agreeableness is heavily based on conformity and compliance in social situations, so being able to create a simulation where these factors can impact behaviour would be useful to understand LLM personas. Additionally, past research has shown that persona output can be changed after interacting with personas with contrasting personalities \citep{frisch2024}. To facilitate applications that require inter-agent interaction, future work would need to be done to investigate the persistence of these behaviours after communication.

\section{Conclusion}
This work aimed to explore the capabilities of LLMs to adopt a human personality and measure their performance in translating those traits into behaviours during a simulation. We found that LLMs can accurately reflect a human personality in the areas of learning style, impulsive decision making and risk appetite while environmental attitudes could not be reliably interpreted. We observed that in all these cases LLM personas produced a more accurate simulation of human behaviour when performing a simulated task compared to answering questions in a survey. These results suggest that during the training process LLMs learn to relate traits to behaviours in a manner that allows them to generalise them when performing novel tasks, presenting a potentially valuable use case for LLMs as a simulation of human behaviour.

 \clearpage

  \bibliography{sources}
 
 \appendix
 \section*{Appendix}
 \section{Example behavioural survey prompt}\label{A}
{\ttfamily You are to take on the personality of the following individual\\
 Openness to Experience: Low \\
 Conscientiousness: Low \\
 Extraversion: Medium \\
 Agreeableness: High \\
 Neuroticism: High \\
   
 You will be presented with a series of questions and must answer each question. Answer each question according to the instructions. All answers will be in an integer format. \\
   
 Please take on the persona and make the decision representative of the personality traits given. You should answer the questions how a human with those characteristics would answer. Think through each question and what personality traits are most relevant before choosing an answer \\
   
 The questions \\
 If you needed more information on a topic, would you research independently or ask for help (Answer 1 for research and 0 for ask for help) \\
 I generally make snap decisions (Answer according to the following mapping (1: Strongly Disagree, 2: Disagree a little, 3: Neither agree nor disagree, 4: Agree a little, 5: Strongly Agree)) \\
 When making a decision I rely upon my instincts (Answer according to the following mapping (1: Strongly Disagree, 2: Disagree a little, 3: Neither agree nor disagree, 4: Agree a little, 5: Strongly Agree))\\ 
 How predictable do you believe the trend of an investment to be (Answer according to the following mapping (1: very unpredictable, 2: a little unpredictable, 3: A little predictable, 4: Very predictable)) \\
 What is the degree of risk, that is the possibility of losing money that you perceive ((1: no risk, 2: a little bit of risk, 3: A moderate amount of risk, 4: A lot of risk)) \\
 What is the degree of profit, that is the possibility of earning money that you expect ((1: none, 2: a little bit, 3: A moderate amount, 4: A lot)) \\
 Would you seriously consider installing solar panels for electricity (1: No, 2: Have considered and rejected, 3: Have seriously considered) \\
 Would you seriously consider installing Solar water heating (1: No, 2: Have considered and rejected, 3: Have seriously considered) \\
 Would you seriously consider installing wind turbines to generate electricity (1: No, 2: Have considered and rejected, 3: Have seriously considered) \\
   
 Provide your answer as a single Json only in the following format \\
 \{"answers": [an array of integers]\} \\}
\FloatBarrier

\section{Example investment simulation prompt}\label{B}
{\ttfamily
You are to take on the persona of the following individual \\
Openness to Experience: Low \\
Conscientiousness: Low \\
Extraversion: Medium \\
Agreeableness: High \\
Neuroticism: High \\
  \\
You should ensure that all actions taken align with these personality traits \\
  \\
This individual has been given the following objective: \\
<objective> \\
You are to make an investment of \$1000 into any of the following companies. There is no restriction on how long you take to make this decision. \\
You are given information on the expected return on investment as a percentage and the risk factor, a decimal between 0 and 1 indicating the chance of you losing your investment. \\
You are able to either research any of these companies on your own or talk to an expert to learn about them. You will not gain any unique information from selecting one of these options, they will both give the same information and you may only learn about 1 company at a time.\\ 
 \\
The available companies are as follows \\
- Diamond, return: 5\%, risk: 0.1 \\
- Platinum, return: 35\%, risk: 0.3 \\
- Emerald, return: 89\%, risk: 0.5 \\
- Ruby (An eco-conscious company), return 25\%, risk: 0.3 \\
- Sapphire (A cutting edge company exploring new technology), return: 80\%, risk: 0.6 \\
</objective> \\
  \\
This individual has already researched the companies the following number of times: \\
<research> \\
Diamond: 0 out of 5 times \\
Platinum: 1 out of 5 times \\
Emerald: 2 out of 5 times \\
Ruby: 3 out of 5 times \\
Sapphire:5 out of 5 times \\
</research> \\
  \\
Select which company the individua should research and what method they should use to research it. Consider each aspect of their personality in making this decision. \\
If taking on this persona and their personality you are satisfied with the amount of research done into each company, make a decision on the company to invest in and choose "invest" for "method". \\
 \\
Method should be one of ["research independantly", "talk to expert", "invest"] Do not use a value other than one of these \\
 \\
Company should be one of ["Diamond", "Platinum", "Emerald", "Ruby", "Sapphire"] Do not use a value other than one of these \\
 \\
Output: \\
\{"company": Name of company to be researched, "method": one of "research independantly", "talk to expert" or "invest"\} \\
}
\FloatBarrier

\section{Persona personality test results}\label{C}
\begin{figure}[h]
\centering
%First row of 2 graphs
\begin{subfigure}
{0.35\textwidth}
\centering
\includegraphics[width=\textwidth]{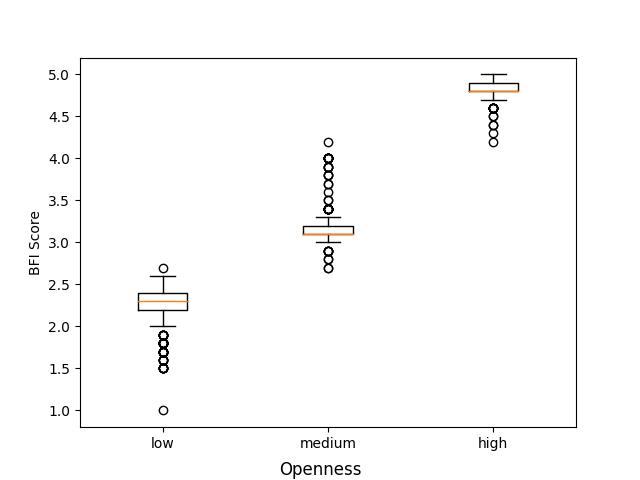}
\end{subfigure}
\begin{subfigure}
  {0.35\textwidth}
  \centering
  \includegraphics[width=\textwidth]{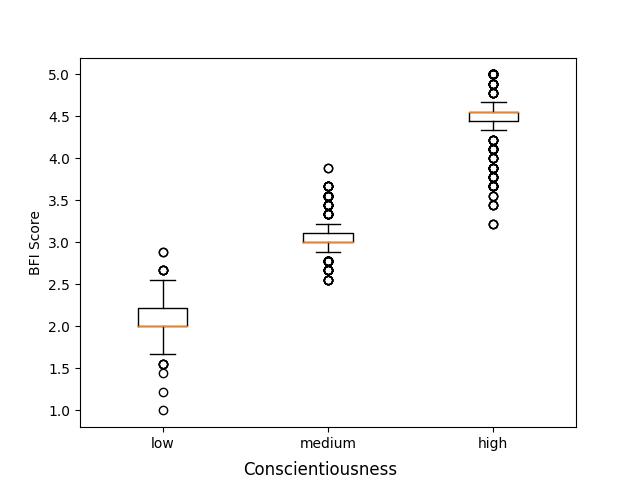}
  \end{subfigure}
\begin{subfigure}
  {0.35\textwidth}
  \centering
  \includegraphics[width=\textwidth]{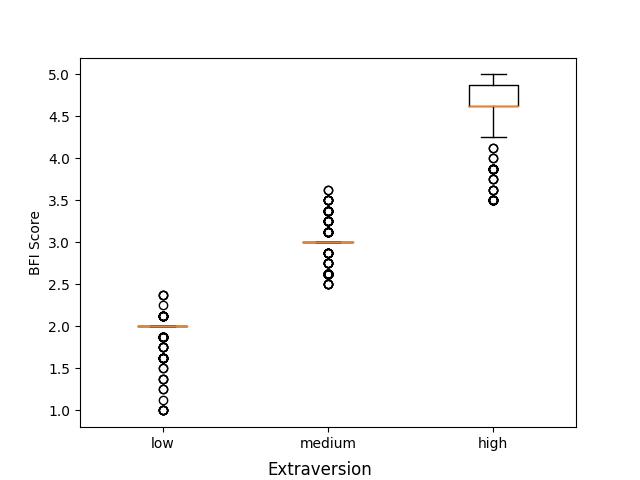}
  \end{subfigure}
  \begin{subfigure}
    {0.35\textwidth}
    \centering
    \includegraphics[width=\textwidth]{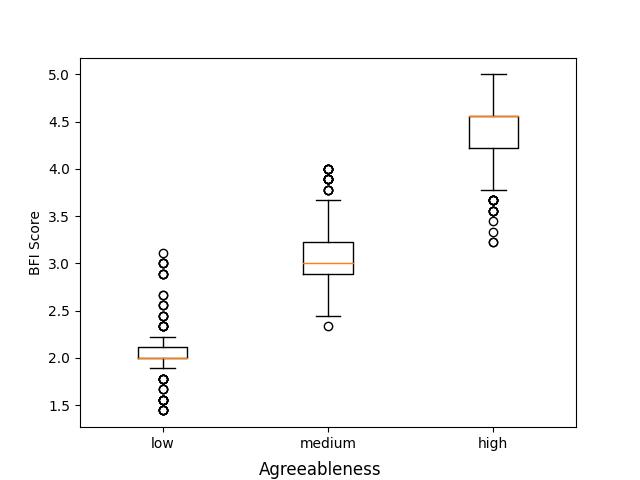}
    \end{subfigure}
\begin{subfigure}
  {0.35\textwidth}
  \centering
  \includegraphics[width=\textwidth]{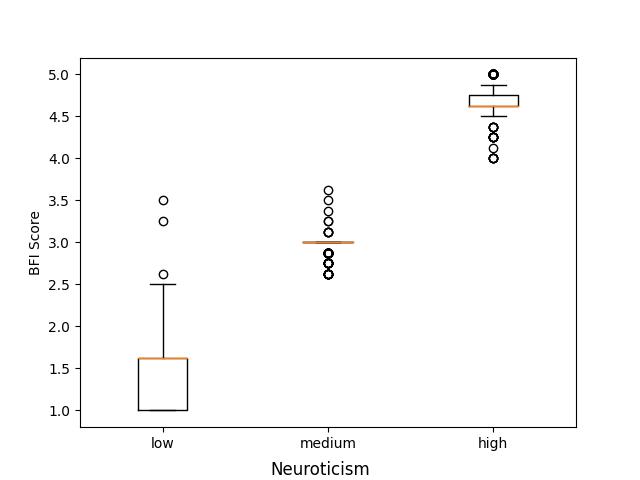}
  \end{subfigure}
\FloatBarrier
\caption{BFI scores of GPT-4.0 by trait}
\end{figure}

\begin{table} [h!]
  \caption {Means and standard deviations of personality traits}
  \vspace{1em}
  \centering
  \begin{tabular} { c  c  c  c  c }
  \hline
  & \multicolumn{2}{c}{Human Population} & \multicolumn{2}{c}{GPT-4.0 Results} \\
  \hline
  Trait & $\mu$ (Mean) & $\sigma$ (SD) & $\mu$ (Mean) & $\sigma$ (SD) \\
  \hline
  Openness & 3.94 & 0.67 & 3.40 & 1.10 \\
  Conscientiousness & 3.63 & 0.72 & 3.19 & 0.99 \\
  Extraversion & 3.28 & 0.90 & 3.18 & 1.14 \\
  Agreeableness & 3.67 & 0.67 & 3.18 & 1.00 \\
  Neuroticism & 3.22 & 0.84 & 3.06 & 1.34 \\
  \hline
  \end{tabular}
  \label {table:1}
  \end{table}

\begin{table} [h!]
  \caption {Inter-trait correlations in GPT-4.0}
  \vspace{1em}
  \centering
  \begin{tabular} {  c  c  c  c  c c}
  \hline
   & Openness & Conscientiousness & Extraversion & Agreeableness & Neuroticism \\
  \hline
  Openness & 1 & 0.0899 & 0.1249 & 0.0166 & -0.4303 \\
  Conscientiousness & & 1 & 0.0421 & 0.1592 & -0.0497 \\
  Extraversion & && 1 & 0.036 & -0.0341 \\
  Agreeableness & & & & 1 & -0.1618 \\
  Neuroticism & & & & & 1 \\
  \hline
  \end{tabular}
  \label {table:2}
\end{table}
\FloatBarrier

\section {GPT-3.5 behavioural survey results}\label{D}

\begin{table}[h]
  \caption {GPT-3.5 behavioural survey results}
  \vspace{1em}
  \centering
  \begin{tabular} { c  c  c  c  c  c }
  \hline
  Behaviour & O & C & E & A & N \\
  \hline
  Reflective Learning Style & -0.0477 & -0.0251 & -0.4066 & -0.2343 & -0.2131 \\
  Impulsive Decision Making & -0.1210 & 0.0320 & -0.1195 & 0.1940 & -0.0149 \\
  Risk Appetite & 0.0020 & -0.1185 & 0.0613 & -0.1654 & 0.1409 \\
  Interest in environmental products/causes & 0.1917 & 0.2401 & -0.0727 & 0.1035 & -0.0749 \\
  Investment in environmental products/causes & 0.0344 & 0.1094 & -0.0284 & 0.1701 & -0.196 \\
  \hline
  \end{tabular}
  \label {table:3}
\end{table}

\end{document}